# Reionization in HESTIA: studying reionization in the LG through zoom simulations

David Attard[1,2]★ Luke Conaboy[3] Noam Libeskind,[4] Sergey Pillipenko,[5] Keri Dixon[6] and Ilian T. Iliev[1]

[1]*Department of Physics & Astronomy, University of Sussex, Brighton BN1 9QH, UK*
[2]*DISCnet Centre for Doctoral Training, University of Sussex, Brighton BN1 9QH, UK*
[3]*School of Physics and Astronomy, The University of Nottingham, University Park, Nottingham NG7 2RD, UK*
[4]*Leibniz-Institut für Astrophysik Potsdam (AIP), An der Sternwarte 16, D-14482 Potsdam, Germany*
[5]*Astro Space Center, P.N. Lebedev Physical Institute, Profsojuznaya 84/32, Moscow 117997, Russia*
[6]*Department of Physics, Geosciences, and Astronomy, Eastern Kentucky University, 521 Lancaster Ave., Richmond, KY 40475, USA*

## ABSTRACT

While cosmic reionization has been broadly constrained by global observables, the interplay between internal sources [Milky Way (MW), M31, and their satellites] and external ionization fronts remains poorly understood in a realistic Local Group (LG) context. To address this issue, we perform radiative transfer post-processing on the original HESTIA LG constrained simulation. We calibrate our source models using a uniform $1024^3$ particle, dark matter-only, HESTIA simulation coupled with a subgrid collapse fraction model to match the global reionization observables. These source models are then applied to the HESTIA zoom-in simulations, which consist of a $4096^3$ particle effective resolution in the zoom region centred on the MW and M31 haloes, which resolves haloes down to $10^8$ M$_\odot$. We find that in all scenarios, reionization within the LG proceeds in an inside-out manner with the progenitors of the MW and M31 having 50 per cent of their material ionized by $z \approx 9$–8.6, significantly earlier than the global mid-point at $z \approx 7$–7.7, noting that external fronts from large-scale structure play a negligible role, even under the most permissive feedback model. We further show that present-day satellite galaxies exhibit only a weak correlation between their reionization redshift and their present-day radial distance from their host halo, with somewhat tighter trends around M31 than the MW. Finally, we find that present-day satellites whose assembly preceded the reionization of most of their $z = 0$ material are systematically more massive today, suggesting that the oldest stellar populations preferentially reside in the most massive $z = 0$ subhaloes.

## 1 INTRODUCTION

The epoch of reionization (EoR) represents one of the last global transitions of the Universe whereby it went from cold and largely neutral to almost completely ionized and hot (for review see e.g. S. Zaroubi 2013; N. Y. Gnedin & P. Madau 2022). This transition occurred over an extended period of time, roughly between redshifts 15 and 6 (R. Barkana & A. Loeb 2001; X. Fan et al. 2002; I. D. McGreer, A. Mesinger & V. D'Odorico 2015; C. A. Mason et al. 2019; F. Wang et al. 2020; M. Nakane et al. 2024). It marked the end of the cosmic 'dark ages' and set the stage for subsequent galaxy formation and evolution. During this period, the emergence of Population III, metal-free stars, dwarf galaxies, and later quasars generated copious amounts of ultraviolet (UV) radiation, which created expanding ionized 'bubbles' that eventually percolated to fill the entire intergalactic medium (IGM; S. L. Finkelstein 2016; S. Hassan et al. 2018; J. Rosdahl et al. 2018; M. Trebitsch, M. Volonteri & Y. Dubois 2020; H. Atek et al. 2024).

Despite substantial progress in both theory and observation, the details of reionization remain subject to significant uncertainties. Key questions include: What were the dominant ionizing sources? How did radiative feedback and the inhomogeneous density of the IGM shape the progression of reionization? And crucially, how did local environmental factors influence the timeline and morphology of the reionization process? Recent studies have emphasized that, although the global reionization history can be broadly constrained by cosmic microwave background (CMB) measurements (Planck Collaboration VI 2020) and quasar absorption spectra (L. C. Keating et al. 2024), the spatial and temporal variability in the ionization field likely leaves important imprints on the properties of the galaxies we see today (J. S. W. Lewis et al. 2022).

The Local Group (LG), comprising the Milky Way (MW), Andromeda (M31), and their satellite galaxies, offers a unique laboratory to study these open questions. The Constrained Local Uni-versE Simulations (CLUES)[1] collaboration has been pivotal in providing realistic initial conditions (ICs) for simulations of our local environment (S. Gottloeber, Y. Hoffman & G. Yepes 2010; N. I. Libeskind et al. 2010; E. Carlesi et al. 2016). By combining observational data with sophisticated modelling techniques, CLUES produces constrained realizations of the local density and velocity fields that, when evolved, reproduce the local Universe at the present time. These tailored ICs underpin the HESTIA simulation suite (N. I. Libeskind et al. 2020), and enable us to reproduce the complex structure of the LG, including the MW–M31 pair and their satellites, within their proper large-scale environment, including nearby galaxy clusters, voids, and other structures like the Great Attractor. Constrained simulations such as those from the HESTIA suite allow us to model the LG with high resolution in a realistic cosmic context, enabling detailed exploration of how feedback processes and local ionizing sources shaped reionization within our immediate neighbourhood.

★ E-mail: da500@sussex.ac.uk

**Table 1.** Summary of the three $N$-body simulations used in this study ran with the codes CUBEP$^3$M (J. Harnois-Deraps et al. 2013) [1] and RAMSES (R. Teyssier 2002) [2].

| Simulation | Resolution | Purpose | Code | $m_{\mathrm{DM}}$ ($\mathrm{M_{\odot}}$) |
|---|---|---|---|---|
| S1 | $4096^3$ | Subgrid model statistics | [1] | $2 \times 10^6$ |
| HESTIA-1024 | $1024^3$ | RT calibration | [2] | $1.336 \times 10^8$ |
| HESTIA-ZOOM | $4096^3$ (in zoom region) | LG RT study | [2] | $2 \times 10^6$ (zoom region) |

A key inquiry is whether reionization in the LG proceeded predominantly in an 'inside-out' fashion, where the central, massive galaxies reionized their own surroundings prior to the arrival of external ionizing fronts, or whether external sources contributed significantly to the ionization process. Likewise, the relationship between the reionization redshift ($z_{\mathrm{reion}}$) and the present-day properties of satellite galaxies remains poorly constrained, with some studies indicating reionization relationships between high-mass satellites and their hosts (T. Y. Li et al. 2014; P. Ocvirk et al. 2014), while others reveal a broad scatter that may reflect varying degrees of external influence (K. L. Dixon et al. 2018).

Moreover, both radiative feedback, which comes from ionization and photoheating by ionizing photons, and mechanical feedback, arising from processes such as supernova explosions and stellar winds, may suppress star formation in low-mass haloes, thereby influencing both the local photon budget and the timing of reionization. The high-mass atomic cooling haloes (HMACHs, with virial temperatures $>10^9$ K) are largely unaffected by radiative feedback since the gas in them cools efficiently to form stars. In contrast, the shallow gravitational potentials of low-mass atomic cooling haloes (LMACHs, with virial temperatures below $\sim10^4$ K and virial masses of $\sim10^8$–$10^9$ $\mathrm{M_{\odot}}$) render them particularly susceptible to photoheating (P. R. Shapiro, M. L. Giroux & A. Babul 1994). The UV photons from the early luminous sources heat the gas within these haloes, increasing its pressure and raising the local Jeans mass. As a result, gas accretion was inhibited, and star formation was quenched. This mechanism is thought to be an important factor behind the 'missing satellites' problem in the LG, where many predicted subhaloes remain dark (J. S. Bullock, A. V. Kravtsov & D. H. Weinberg 2000; M. T. Busha et al. 2011; M. Jung et al. 2024). Understanding the interplay between radiative feedback and halo mass is therefore essential not only for modelling reionization but also for explaining the present-day population of satellite galaxies and the existence of dark haloes.

In this paper, we employ detailed radiative transfer (RT) simulations applied to the original HESTIA zoom runs to investigate the reionization history of the LG. Our approach combines a $1024^3$ dark matter-only (DMO) simulation with a subgrid model employed to account for the contribution of unresolved haloes. We consider several ionizing source models, summarized below, with source efficiencies carefully calibrated against the available observational constraints. We examine how different feedback prescriptions affect the timing and topology of reionization, and explore the correlation between present-day halo mass and reionization redshift. Our study not only tests the prevailing inside-out reionization paradigm but also addresses several open questions specific to the LG: Are the MW and M31 truly isolated in their reionization histories, or do external ionizing photons play a non-negligible role? How does the diversity in reionization times among satellites relate to their current mass and spatial distribution? And what constraints can we place on the efficiency of feedback processes, in particular low-mass haloes, which may lead to the suppression of star formation and the emergence of dark haloes?

The remainder of the paper is organized as follows. Section 2 describes our simulation setup, subgrid modelling, and RT methodology. Section 3 presents our analysis of the reionization history in the LG, including comparisons between different feedback models. Section 4 discusses the implications of our findings, summarizes our conclusions, and suggests directions for future work.

## 2 METHODOLOGY

### 2.1 $N$-body simulations

In this study, we performed two $N$-body DMO simulations (HESTIA-1024 and HESTIA-ZOOM) and also used a high-resolution simulation presented in Schwandt et al. (in preparation) (referred to as S1 hereafter), which was used to calibrate our collapse fraction subgrid model described in Section 2.1.1. Table 1 summarizes the DMO simulations used.

#### 2.1.1 *Initial conditions and simulation setup*

The main aim of this paper is to study reionization in the context of the local Universe. To do this, we use ICs based on the original HESTIA suite (N. I. Libeskind et al. 2020). We regenerated the ICs in GRAFIC (RAMSES-readable) format, using the original white noise fields, with the public GINNUNGAGAP code.[2] Finally, we generate refinement masks for the $4096^3$ effective zoom simulation using a modified version of MUSIC (O. Hahn & T. Abel 2011). This process is similar to that used in N. I. Libeskind et al. (2020), where we track all the particles within a $5\,h^{-1}$ Mpc sphere of the LG back to the initial redshift, and then refine the resulting Lagrangian volume.

[1] https://www.clues-project.org/cms/

[2] github.com/ginnungagapgroup/ginnungagap

We initialize our $N$-body DMO simulations using the ICs from the 09_18 realization of the HESTIA simulation suite, all consisting of a $L_{\rm box} = 100\,h^{-1}$ cMpc box. These ICs are specifically designed to reproduce the correct LG cosmography, yielding an MW and M31 halo pair, a Local Void, and a Virgo Cluster consistent with observations. All simulations assume the background cosmology used in HESTIA, with parameters $h = 0.677$, $\sigma_8 = 0.83$, $\Omega_\Lambda = 0.682$, and $\Omega_{\rm M} = 0.318$ consistent with Planck Collaboration XVI (2014).

Our two simulations were evolved from $z = 99$ down to $z = 0$ using the RAMSES code R. Teyssier (2002). The first simulation is a uniform-resolution $N$-body DMO run consisting of $1024^3$ particles, used to calibrate our RT source models and to apply the mock haloes predicted by the subgrid model. The second is a DMO zoom simulation with a base resolution of $256^3$ dark matter (DM) particles, featuring a high-resolution zoom region; a sphere of radius $5\,h^{-1}$ Mpc centred on the LG at $z = 0$, with an effective resolution of $4096^3$ DM particles. We employ adaptive mesh refinement, and allow the grid to refine up to $\ell_{\rm max} = 18$ for both runs, corresponding to a minimum comoving cell size of $381\,h^{-1}$ cpc. Refinement occurs when there are more than eight DM particles in a grid cell. We use the standard refinement scheme in RAMSES, where higher levels of refinement are 'released' as the simulation progresses.

Because reionization in our Universe occurs over large, inhomogeneous scales, often spanning tens of Mpc (I. T. Iliev et al. 2014), zoom simulations face challenges in RT studies. In these simulations, only the haloes within the limited zoom region are resolved, which restricts the ability of ionization fronts from outside the region to be accurately captured. However, previous studies (I. T. Iliev et al. 2011; K. L. Dixon et al. 2018; P. Ocvirk et al. 2020) indicate that, for the LG, reionization largely follows an inside-out progression, suggesting that the zoom approach may still be appropriate for studying reionization in the context of the LG haloes.

We use the HESTIA-1024 simulation to calibrate the ionizing source models employed in our RT simulations with global observational constraints of the EoR, as well as to verify that in the *09_18* HESTIA realization, the LG is being internally ionized and hence zoom simulations can be used to approximate the radiative field within the LG. However, the HESTIA-1024 simulation is only able to resolve (up to 20 bound particles) haloes $>10^{9.4}$ M$_\odot$, which means that it would miss any ionizing contributions from haloes having a mass in the range $10^8 <$ M$_\odot < 10^{9.4}$. These haloes are all within the atomic cooling limit and thus would be able to cool and form stars during the EoR.

To model the contribution from such haloes, we couple our HESTIA-1024 DMO simulation with the instantaneous halo-bias subgrid model presented in A. Nasirudin, I. T. Iliev & K. Ahn (2020). This subgrid model is based on the correlation between the local overdensity ($\delta$) and the collapse fraction ($f_{\rm coll}$) using a high-resolution simulation, in our case S1, where $f_{\rm coll}$ represents the fraction of mass that collapses into haloes. S1 is a simulation consisting of a $100\,h^{-1}$ cMpc box, containing $4096^3$ DM particles and thus is able to resolve haloes with masses down to $10^8$ M$_\odot$ (comprising of $>50$ particles).

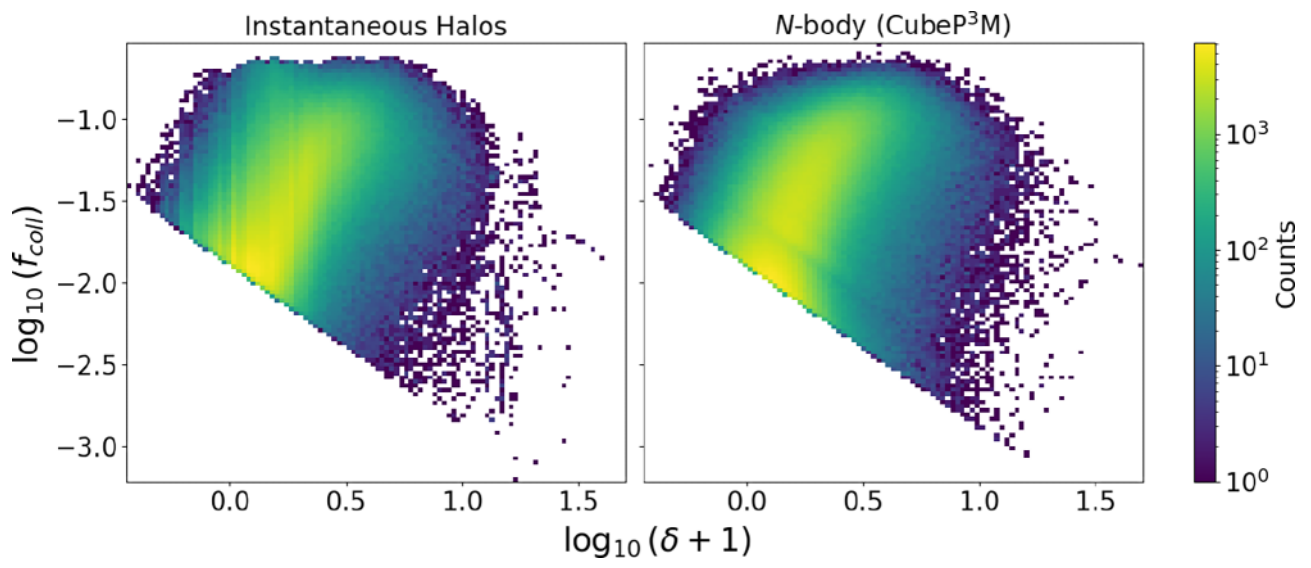

**Figure 1.** Collapsed fraction of LMACHs ($f_{\rm coll,8:9}$) per cell as a function of cell overdensity ($\log_{10}(1+\delta)$) at $z = 8.899$ for a simulation box of $100\,{\rm Mpc}\,h^{-1}$ and a $256^3$ grid ($0.39\,h^{-1}$ Mpc cell size). The panels display instantaneous mock haloes (left) and $N$-body halo data (right). Data points are binned into a two-dimensional grid of $\log_{10}(1+\delta)$ and $\log_{10}(f_{\rm coll,8:9})$, with the colour map representing the number of data points in each bin.

In our study, we divide the ionizing sources into two different categories: HMACHs and LMACHs. We define LMACHs as those haloes with a halo mass $10^8 < M_{\rm halo} < 10^9$ M$_\odot$, and HMACHs as those haloes having a halo mass $>10^9$ M$_\odot$. This value of halo mass was specifically chosen such that haloes with mass $<10^9$ M$_\odot$ would be below the Jeans mass when in ionized regions and thus would be inefficient at accreting gas/forming stars when calculating the emissivity of ionizing photons (K. L. Dixon et al. 2018). Haloes below this threshold, known as minihaloes, are unable to cool efficiently via atomic processes. Although minihaloes may initiate reionization by hosting very metal-poor stars that may have a non-negligible contribution during the early stages (e.g. K. Ahn et al. 2009, 2012; K. T. Abe 2022), we exclude them from our analysis because our focus is on the later stages of reionization, when these haloes are completely suppressed.

To implement the subgrid model, we smooth the density fields of the $1024^3$ DMO and the S1 $N$-body simulation, and coarsen them to a $256^3$ grid, matching the resolution of our RT simulations. The model is then used to predict mock haloes for all the LMACHs in the HESTIA-1024 simulation, as well as to model the unresolved HMACH population ($10^9 < M_{\rm halo} < 10^{9.4}$ M$_\odot$) in the same simulation. For a more detailed explanation of the source models, refer to Section 2.3.

Fig. 1 presents the distribution of the collapsed fractions of haloes with masses in the range $10^8$–$10^9$ M$_\odot$ ($f_{\rm coll}$, 8:9) at $z = 8.899$, comparing the direct $N$-body simulation using S1 data (right panel) with the instantaneous mock haloes applied to the HESTIA-1024 DMO simulation (left panel).

## 2.2 The Local Group

The AHF code (S. R. Knollmann & A. Knebe 2009) was used to identify the haloes in our simulation, where only those haloes with at least 20 bound DM particles were kept for further analysis. MERGERTREE (a tool included in the AHF package) was used to construct merger trees of the identified AHF haloes.

To identify the MW and M31 haloes in our simulations, we apply the same selection criterion employed in N. I. Libeskind et al. (2020), mainly that the LG pair needs to have:

(i) a halo mass of MW and M31 in the range $8 \times 10^{11} < M_{\rm halo}\,[{\rm M}_\odot] < 3 \times 10^{12}$;
(ii) a separation of $0.5 < d_{\rm sep}\,[{\rm Mpc}] < 1.2$;

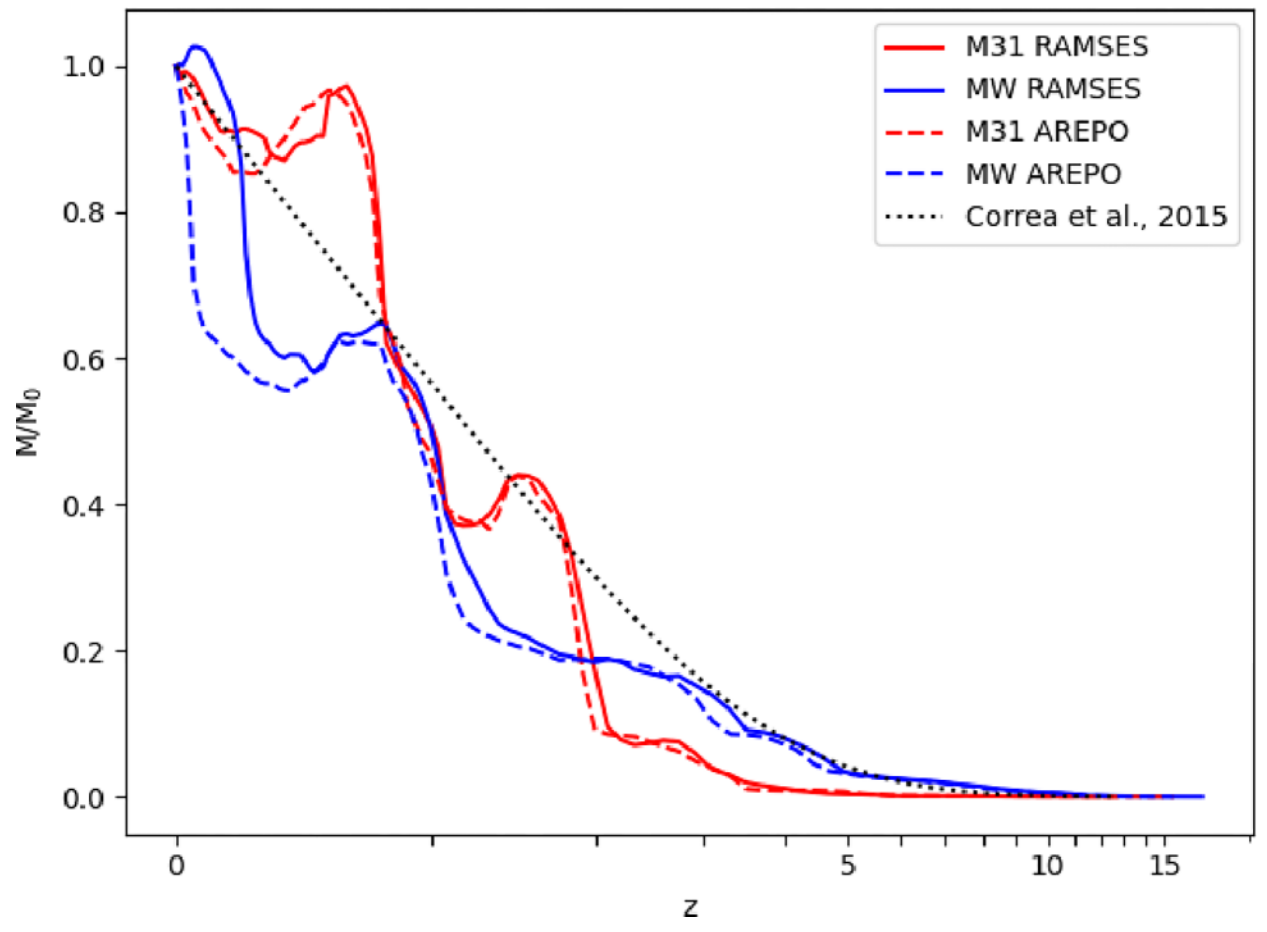

**Figure 2.** The MAH of the largest progenitor halo for the MW halo (blue) and M31 (red), when the constrained ICs were evolved using RAMSES (solid) and AREPO (dashed). In black, we also show the semi-analytical relation from C. A. Correa et al. (2015).

(iii) isolation, such that there is no third halo more massive than the MW within 2 Mpc of the LG pair;
(iv) a halo mass ratio of $0.5 < M_{\rm MW}/M_{\rm M31} < 1$;
(v) be infalling (i.e. $v_{\rm rad} < 0$).

Using the criteria outlined above, we identify the MW and M31 haloes. We then confirm that their mass accretion histories (MAHs), as traced in the RAMSES run, are consistent with those derived from the original HESTIA simulations performed with AREPO (Fig. 2). These MAHs were computed by linking the most massive progenitors across simulation snapshots using the halo merger tree. In our zoom-DMO simulation, the MW halo had a halo mass of $1.889 \times 10^{12}\,{\rm M}_\odot$ at $z = 0$, whilst the M31 halo had a halo mass of $2.152 \times 10^{12}\,{\rm M}_\odot$. The haloes are both infalling with a relative radial velocity of $103.25\ {\rm km\,s}^{-1}$ and are separated by a distance of 729.573 kpc. This LG replica is in good agreement with observations of the LG (A. W. McConnachie 2012; J. D. Diaz et al. 2014; P. R. Kafle et al. 2014; L. L. Watkins et al. 2019).

## 2.3 Radiative transfer and source models

There remains significant uncertainty in both the ionizing photon efficiencies of galaxies and the impact of radiative feedback during reionization. To explore these uncertainties, we consider a suite of RT simulations using PYC $^2$ RAY (P. Hirling et al. 2024), a code that couples RT with non-equilibrium chemistry. All simulations are performed on a $256^3$ grid, the same resolution used in our subgrid model, and are calibrated to available observational data, as explained further in Section 3.1. We vary the models by implementing two different source models along with three distinct sets of source efficiencies. This approach allows us to compare simulations with differing source suppression models and efficiencies, and to assess whether these differences produce distinguishable signatures in the reionization history.

### *2.3.1 RT methodology*

To model the progression of reionization, we employ the PYC $^2$ RAY code, an updated, GPU-accelerated version of the well-established C $^2$ RAY code. PYC $^2$ RAY couples the time-dependent RT of ionizing photons with non-equilibrium chemistry, tracking the ionization fractions and temperature of hydrogen (G. Mellema et al. 2006; P. Hirling et al. 2024).

The code utilizes the ASORA (Accelerated Short-characteristics Octahedral Ray-tracing) method to solve the RT equation. A key feature of this method is that it is explicitly photon-conserving. It traces a limited number of rays from each source to every receiving cell using the short-characteristics technique, calculating the column density of neutral gas along the path. By ensuring explicit photon conservation, the method remains stable and accurate even when dealing with large, optically thick cells and regions. This code has been benchmarked against a number of exact analytical solutions (P. Hirling et al. 2024) and directly compared to a standardized set of benchmark problems using a number of independent RT methods (I. T. Iliev et al. 2006b, 2009).

The non-equilibrium chemistry solver updates the ionization state and temperature of the gas based on the calculated photoionization and heating rates, as well as collisional ionization and recombination processes.

Following common practice in post-processing RT simulations, we assume that in the IGM the baryonic density traces the underlying DM density field. This density field is mapped onto a $256^3$ regular grid by smoothing the $N$-body particles from the DMO runs using an SPH (smoothed particle hydrodynamics) like technique (as described in Y. Mao et al. 2020), excluding the haloes to focus on the diffuse gas distribution.

### *2.3.2 Source modelling*

The source models adopted here were previously introduced in K. L. Dixon et al. (2016). We define a source as any DM halo identified by AHF and/or implemented by the subgrid model, where we split our sources into LMACHs ($M_{\rm halo} < 10^9\,{\rm M}_\odot$) and HMACHs ($M_{\rm halo} \geq 10^9\,{\rm M}_\odot$). For our $1024^3$ uniform grid DMO simulation, the minimum halo mass resolved (with at least 20 bound particles) is that of $10^{9.4}\,{\rm M}_\odot$, thus haloes with masses $10^8 < M_{\rm halo} < 10^{9.4}\,{\rm M}_\odot$ had to be implemented using subgrid modelling as explained in Section 2.1.1. In the zoom simulations, all haloes with $M_{\rm halo} > 10^8\,{\rm M}_\odot$ are well resolved. In this work, we do not consider the effects of minihaloes with masses less than $10^8\,{\rm M}_\odot$. Although these structures may have played a role in ionization during the early EoR, their influence in the later stages is expected to be negligible. This is due to the dissociation of molecular hydrogen by UV background radiation from early luminous sources, making their contribution insignificant compared to more massive atomic cooling haloes (K. Ahn et al. 2009).

In the models presented in K. L. Dixon et al. (2016), the ionizing photon emissivity of a source, $\dot{N}_\gamma$, is dependent on the source efficiency, $g_\gamma$, and the mass of the halo, $M_{\rm halo}$, following the relation

$$\dot{N}_\gamma = g_\gamma \frac{M_{\rm halo}\,\Omega_{\rm b}}{m_{\rm p}\,\Delta t_{\rm s}\,\Omega_{\rm M}}, \quad (1)$$

where $\Omega_{\rm b}$ is the baryon density parameter, $m_{\rm p}$ is the mass of a proton, $\Delta t_{\rm s} = 10\,{\rm Myr}$ is the lifetime of the star (set to equal the time between $N$-body snapshots), and $g_\gamma$ is the efficiency factor defined as

$$g_\gamma = f_*\, f_{\rm esc}\, N_{\rm i}, \quad (2)$$

where $f_*$ is the star formation efficiency, $f_{\rm esc}$ is the photon escape fraction, and $N_{\rm i}$ is the number of ionizing photons produced per baryon in stars, which depends on the initial mass function (IMF) and metallicity. Where e.g. $\dot{N}_\gamma \sim 4000$ for a Population II star when assuming a Salpeter IMF, but the values for $f_*$ and $f_{\rm esc}$ remain uncertain, and thus these parameters can be tuned to match the observed constraints discussed in Section 3.1. The two source models used were:

(i) *Full suppression model*. In this model, HMACHs are assigned a source efficiency of $g_{\gamma,\rm HMACH}$. LMACHs are assigned a source efficiency of $g_{\gamma,\rm LMACH}$ in neutral regions, with $g_{\gamma,\rm LMACH} > g_{\gamma,\rm HMACH}$. These early small galaxies can harbour the more massive Population III stars, which would produce more photons. In ionized regions, LMACHs are assumed to be completely suppressed and not able to produce any ionizing photons, due to mechanical or radiative feedback or a combination of both. Such a source model is considered our maximal model, where feedback is very strong.

(ii) *Partial suppression model*. Similar to the full suppression model HMACHs are assigned a source efficiency of $g_{\gamma,\rm HMACH}$ and LMACHs are assigned a source efficiency of $g_{\gamma,\rm LMACH}$ in neutral regions such that $g_{\gamma,\rm LMACH} > g_{\gamma,\rm HMACH}$. However, in ionized regions, LMACHs are only partially suppressed from producing ionized photons and are given the same source efficiency as HMACHs. This model might depict a scenario in which the photoheating of surrounding gas impacts or shuts off any new gas supply, but a residual gas reservoir within the galaxy itself could support some star formation.

To explore the impact of varying suppression models and ionizing efficiencies, we define four primary RT scenarios, each run both in the uniform subgrid simulation and in the high-resolution zoom-in simulation. These runs differ by their choice of suppression model and photon production efficiencies for HMACHs and LMACHs:

(i) Run 1: partial suppression model with $g_{\gamma,\rm HMACH} = 0.7$, $g_{\gamma,\rm LMACH} = 1.5$;
(ii) Run 2: full suppression model with $g_{\gamma,\rm HMACH} = 1.5$, $g_{\gamma,\rm LMACH} = 2$;
(iii) Run 3: partial suppression model with $g_{\gamma,\rm HMACH} = 1.1$, $g_{\gamma,\rm LMACH} = 1.7$;
(iv) Run 4: full suppression model with $g_{\gamma,\rm HMACH} = 1.1$, $g_{\gamma,\rm LMACH} = 1.7$.

Each of these four setups is applied to both simulation types, yielding eight runs in total: `Run1_subgrid`, `Run1_zoom`,..., `Run4_zoom`. We use a naming convention of `Run#_subgrid` or `Run#_zoom` to differentiate between these. Runs 1 and 2 were calibrated to produce the same global reionization history, enabling a direct comparison of the effects of different suppression models. In contrast, Runs 3 and 4 use identical efficiency parameters but different suppression models, allowing us to isolate the impact of feedback strength. This design provides insight into whether global reionization observables can differentiate between scenarios with similar histories but different underlying physics. All our RT simulations utilize a $256^3$ grid.

After verifying that our simulations produced a physically consistent reionization scenario in the $100\,h^{-1}$ Mpc box using the subgrid model coupled RT runs, the same RT scenarios were applied to the zoom simulation. This allowed us to gain a more detailed understanding of how reionization affects satellite galaxies, which remain unresolved in the $1024^3$ DMO simulation. It is important to note that in the zoomed simulation, most haloes outside the zoom region are not resolved, and as a result, their ionizing radiation cannot be fully captured.

### 2.4 Halo reionization times

Since in this study, we are post-processing RT to our DMO simulations, we cannot directly attribute the reionization time of a specific halo depending on the properties of the gas bound to that halo. We assign the redshift of reionization, $z_{\rm reion}$, to each halo identified at $z = 0$ based on the reionization times of all the DM particles that ultimately reside in the halo, similar to the method used in D. Aubert et al. (2018).

We track the position of each DM particle in a given halo and, at every snapshot ($z_j$), assign each particle ($p_i$) an ionization fraction based on the ionization fraction of the RT cell in which it is found, denoted as $x_{\rm H\,II,\mathit{p_i}}(\boldsymbol{x}, z_j)$. Thus, at each redshift, we can calculate the ionized fraction of each halo by calculating the mass-weighted average:

$$x_{\rm H\,II,\mathit{z_j}} = \frac{\sum_{p_i\in\rm halo} x_{\rm H\,II,\mathit{p_i}}(\boldsymbol{x}, z_j)}{\sum_{p_i\in\rm halo} 1}. \quad (3)$$

We define $z_{\rm reion}$ as the redshift at which $x_{\rm H\,II,\mathit{z_j}}$ exceeds 50 per cent. This method has the advantage over methods that use progenitor positions instead, as it allows us to assign a $z_{\rm reion}$ value to all of the haloes, even those whose progenitors had not yet formed during the EoR.

## 3 RESULTS

### 3.1 Global results

In Fig. 3, we present the globally averaged results from our subgrid model simulations. For the zoom simulations, haloes are identified only within the zoom region, preventing the derivation of global statistics in these cases, hence the reason why these are not present in the figure. By design, all our models are well-calibrated, i.e. show a good agreement with the observational constraints. Notably, we aimed for Run 1 and Run 2 to follow very similar global reionization histories despite employing different suppression models, while Run 3 and Run 4 are relatively more extreme scenarios, roughly marking the earliest and latest reionization that is still consistent with observational constraints. This setup enables a direct comparison of the effects of the global reionization history on the LG under varying assumptions about suppression and the reionization timing. Run 3 finishes reionization the fastest due to the partially suppressed model assumed and also the high photon efficiency rates. Conversely, Run 4 has the slowest reionization history due to the full suppression model, which suppresses all the smaller haloes in ionized regions. Below we summarise the observational constraints used to calibrate our $g_{\gamma,\rm LMACH}$ and $g_{\gamma,\rm HMACH}$.

#### 3.1.1 *Neutral hydrogen fraction*

The top panel of Fig. 3 shows the global volume-weighted neutral hydrogen fraction $x_{\rm H\,II}$ along with a range of observational constraints (M. Ouchi et al. 2010; I. D. McGreer et al. 2015; B. Greig & A. Mesinger 2017; F. B. Davies et al. 2018; C. A. Mason et al. 2018; A. Hoag et al. 2019; I. Jung et al. 2020; D. Ďurovčíková et al. 2024; M. Nakane et al. 2024; H. Umeda et al. 2025). All of our models finish reionization at around $z \sim 6$, although Run 4 still

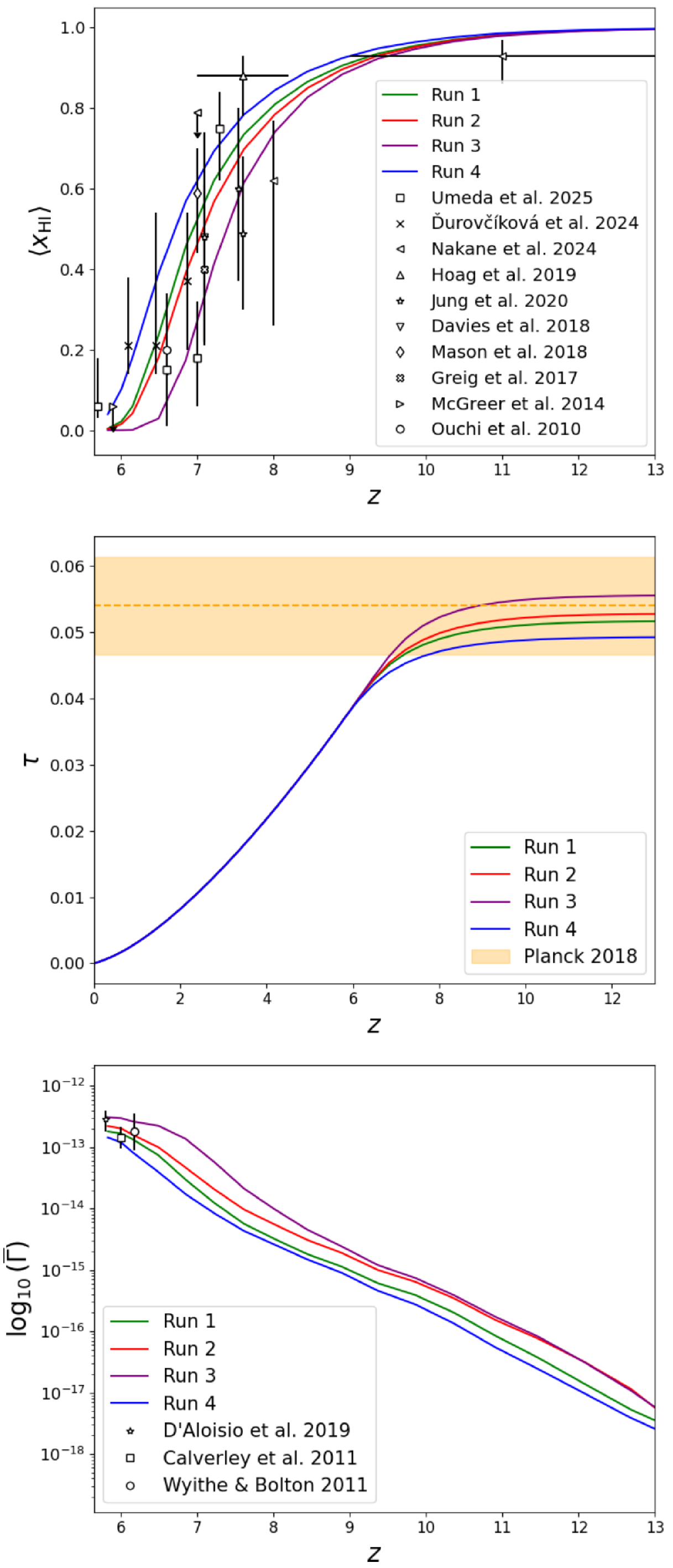

**Figure 3.** The four different runs using the subgrid model simulation. Top: Volume-weighted mean neutral hydrogen fraction compared to observational constraints (M. Ouchi et al. 2010; I. D. McGreer et al. 2015; B. Greig & A. Mesinger 2017; F. B. Davies et al. 2018; C. A. Mason et al. 2018; A. Hoag et al. 2019; I. Jung et al. 2020; D. Ďurovčíková et al. 2024; M. Nakane et al. 2024; H. Umeda et al. 2025). Middle: Thomson optical depth, $\tau$, with *Planck* constraints from Planck Collaboration VI (2020) (yellow). Bottom: Volume-weighted hydrogen photoionization rate compared to constraints (A. P. Calverley et al. 2011; J. S. B. Wyithe & J. S. Bolton 2011; A. D'Aloisio et al. 2019).

has not finished reionization, and fit most of the observational constraints reasonably well.

### *3.1.2 Optical depth to reionization*

The middle panel of Fig. 3 shows the mean integrated Thomson optical depth ($\tau$) as measured from our four different reionization models, where Run 3 and Run 4 exhibit the largest and smallest $\tau$ measurements, respectively, as is expected due to their late/early end of reionization. The plot also shows the $\tau$ measurements made by the Planck Collaboration VI (2020) measured at $\tau = 0.0540 \pm 0.0074$. The value of $\tau$ reflects the influence of free electrons on photons originating from the last-scattering surface of the CMB. It provides a robust, global, and model-independent constraint. All our runs satisfy the Planck Collaboration VI (2020) observational constraint.

### *3.1.3 Photoionization rate*

In the bottom panel of Fig. 3, we show the global photoionization rate for the four different runs using the subgrid DMO simulations versus some selected observational constraints. It can be observed how the evolution of the photoionization rate directly correlates with the reionization histories, with Run 3 having the largest photoionization rates corresponding to its earlier reionization. The increased presence of ionizing photons in the IGM accelerates the reionization process in these scenarios. In contrast, simulations that undergo later reionization show comparatively lower photoionization rates. The photoionization rates in all four runs are consistent with the observations (A. P. Calverley et al. 2011; J. S. B. Wyithe & J. S. Bolton 2011; A. D'Aloisio et al. 2019).

## 3.2 Local results

### *3.2.1 Reionization in the LG region*

The goal of this section is to investigate the reionization history of the LG, specifically the regions surrounding the progenitors of the MW and M31 haloes. We aim to determine whether these regions were reionized internally by their own sources or externally by large-scale ionization fronts, and how this process varies depending on the adopted source and feedback models.

We examine this by creating reionization redshift maps, as follows. For each simulation, we track the ionized fraction of each cell across time. By scanning the simulation outputs from the latest to the earliest redshift, we identify the redshift at which each cell becomes more than 50 per cent ionized, defining this as the cell's reionization redshift. These values are then used to construct 2D projections of the average reionization redshift along the line of sight.

Fig. 4 presents a 1.17 $h^{-1}$ Mpc projection of the resulting reionization map, centred on the location of the M31 main progenitor at $z \sim 9.5$, found by following the merger tree of the most massive progenitor down from $z = 0$. The positions of the MW and M31 progenitor haloes at this redshift are overplotted, with marker sizes proportional to their halo masses as identified in the zoom-in DMO simulation. The white regions indicate cells that remained neutral at the end of the simulation, these correspond to areas outside the zoom region, where no sources were resolved.

The top panels of Fig. 4 show the results from the subgrid-coupled simulations, while the bottom panels display the zoom-in simulations. A clear resemblance between the two sets of simulations can be observed, despite the absence of external ionizing fronts in the zoom-in runs. In both cases, the regions surrounding the MW and M31 progenitors are fully reionized by their own sources. Only the regions that reionize during the very late stages in the subgrid-coupled simulations remain neutral in the zoom-in versions. These are low-density regions without resolved sources, which would have been reionized externally by large-scale I-fronts, e.g. from massive nearby clusters like proto-Virgo that are absent in the zoom simulations (I. T. Iliev et al. 2011; M. Neyer et al. 2024).

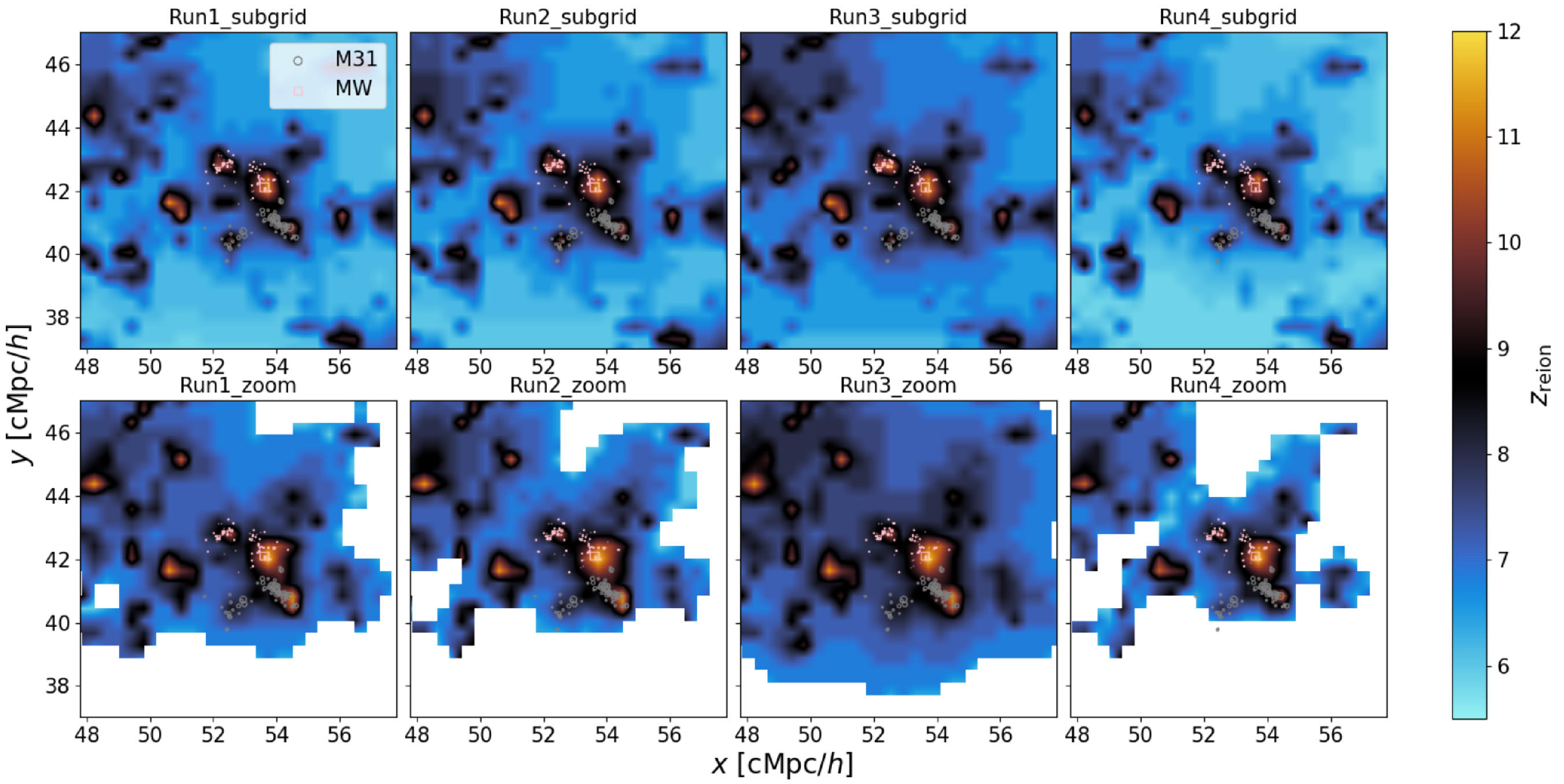

**Figure 4.** A 1.17 $h^{-1}$ Mpc projection of the reionization map centred on the location of the M31 main progenitor at $z = 9.5$. The average redshift of reionization along the projection is shown. The location of the M31 and MW progenitors at $z = 9.5$, as identified through the zoom DMO simulation, is projected onto the reionization map, with the size of the markers being proportional to the halo masses. The white regions mark the cells that are not reionized by the end of the simulations. This is because haloes are only resolved in the zoom region for the zoom simulations.

Comparing different sources and feedback models reveals interesting trends. For example, while Run1_zoom and Run2_zoom predict that reionization ends at roughly the same time, the ionized volume in Run1_zoom is noticeably larger. This is due to the more aggressive source suppression model in Run2_zoom, which limits the ability of smaller sources to contribute to ionization. A similar effect is seen when comparing Run3_zoom and Run4_zoom: although both use the same source efficiencies, the region ionized in the partially suppressed model is significantly larger than in the fully suppressed model.

In all simulations, the more massive haloes begin ionizing their surroundings first. During this early stage, feedback effects are minimal, so the reionization redshifts of these central, high-density regions (appearing red in the map) are relatively unaffected by the suppression model. However, as reionization progresses outward, feedback becomes more significant. This limits the contribution of smaller sources, particularly in the fully suppressed model, where reionization is less extensive.

Despite the differences in ionized volumes, all simulations consistently show that reionization in the LG proceeds in an inside-out fashion without any interference from any incoming ionization fronts from external sources such as Virgo. This suggests that zoom-in simulations are generally sufficient to study the reionization history of the MW, M31, and their satellite systems, as the key physical processes occur within the high-resolution region. This is consistent with I. T. Iliev et al. (2011) and K. L. Dixon et al. (2018), where we found that low-ionizing efficiency models, favoured by current observations, predominantly ionize internally.

### 3.2.2 *The MW and M31 reionization history*

In this section, we examine the reionization history of the Lagrangian mass of the present-day MW and M31 haloes. Fig. 5 presents the mass-weighted ionization fraction of the entire simulation volume in comparison to the MW and M31 haloes across all four runs.

The global reionization history of the full simulation box is derived exclusively from the subgrid model simulations. For the MW (blue) and M31 (red) haloes, we show the reionization history obtained from both the subgrid-coupled simulations (dashed lines) and the zoom simulation (solid lines). The mass-weighted ionization fraction is computed using the particle-based method described in Section 2.4. We extract the ionization history of the MW and M31 haloes in the subgrid-coupled simulation by using the particle positions identified in our zoom simulation.

Although the subgrid-coupled simulations are not used for further analysis, since their primary role was to calibrate the source model, we find that the reionization histories of the MW and M31 haloes are in reasonable agreement with those extracted from the zoom-RT simulations. While an exact match is not expected, given that the subgrid model is a statistical tool designed to approximate the average collapse fraction in a region of a given density, the observed consistency further reinforces its validity.

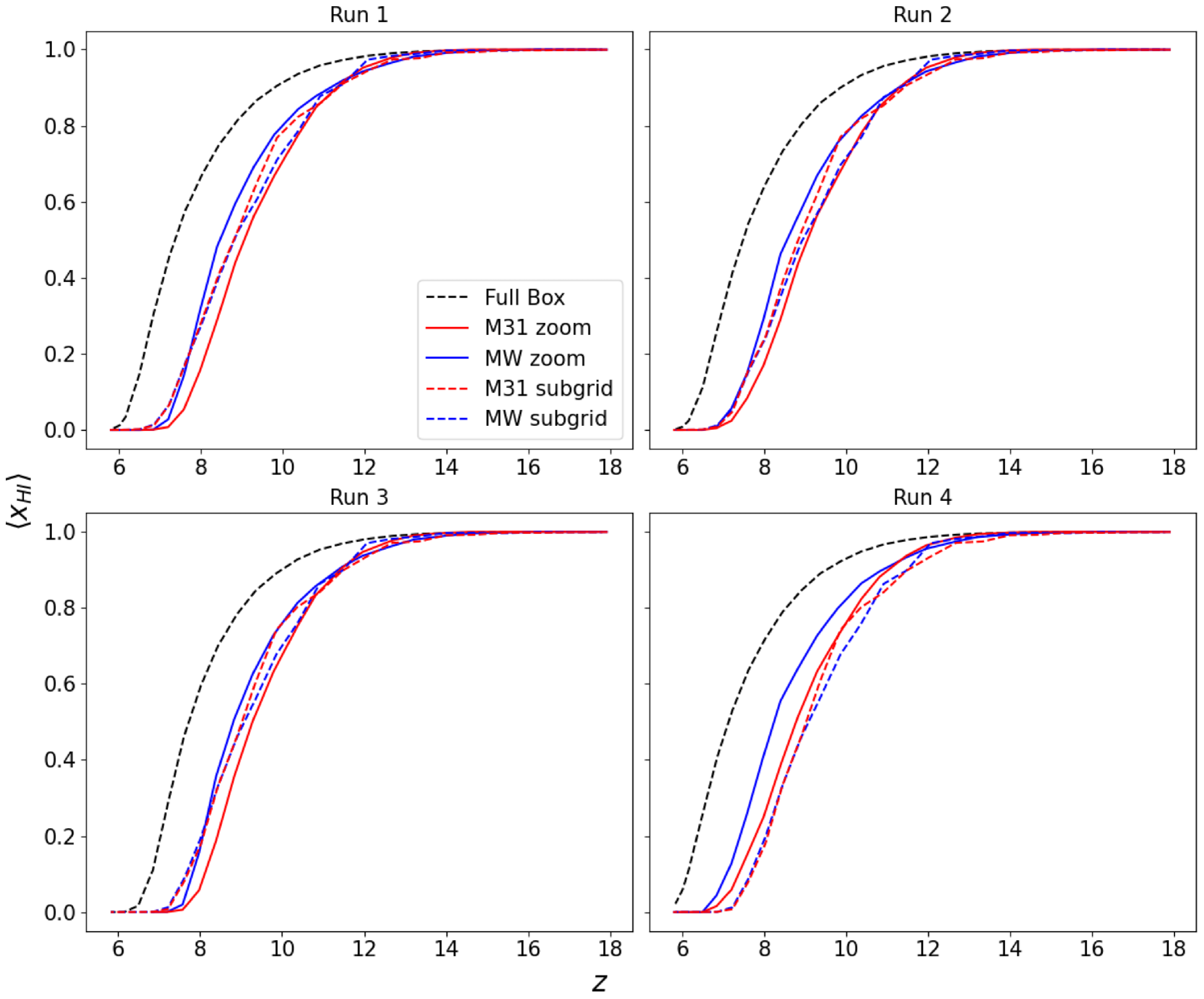

**Figure 5.** The mass-weighted ionization history in the four different runs. The ionization history of the full box in the subgrid-coupled simulations is shown in black. The mass-weighted ionization history of the M31 (red) and MW (blue) haloes is shown, and the results obtained from the subgrid-coupled simulation (dashed lines) and zoom simulations (solid lines) are compared.is required

It can be observed that in all of our runs, both the MW and M31 haloes reionize much earlier than the full box. The full box mass-weighted ionization fraction exceeds the 50 per cent threshold at $z_{\rm reion} = 7.33, 7.43, 7.67,$ and 7.09 for the Run1_subgrid, Run2_subgrid, Run3_subgrid, and Run4_subgrid, respectively. Whilst the redshift of reionization of the M31 and MW haloes in Run1_zoom, Run2_zoom, Run3_zoom, and Run4_zoom are $z_{\rm reion,\,M31} = 9.03, 9.03, 9.25,$ and 8.76, and for the MW, we obtain $z_{\rm reion,\,MW} = 8.87, 8.98, 9.20,$ and 8.65.

When comparing the Run 1 and Run 2 models, it can be noted that despite the slightly earlier reionization of the full box in Run 2, the MW halo still manages to reionize earlier in Run1_zoom as a consequence of the harsher feedback applied in Run2_zoom. The impact of fully suppressing the ionizing output of haloes in ionized regions is further shown when comparing Run 3 and Run 4, where even though the same source efficiencies are used, there is a fairly large difference between the reionization times of the M31 and MW haloes.

Another key observation is that in all our RT runs, it can be observed that the onset for reionization of the MW halo is earlier than the M31 halo. However, in all of our runs $z_{\rm reion,M31} > z_{\rm reion,MW}$. This is because the first progenitor of the MW forms earlier and is more massive at these redshifts than that of M31, however with time the M31 forms more and larger progenitors than the MW halo, in turn producing more ionizing photons, which results in a faster reionization process.

As already discussed earlier, the reionization redshift for the M31 and MW haloes was significantly earlier than the full box. This may be attributed to the fact that the Lagrangian volume of these haloes is in a fairly overdense region (I. T. Iliev et al. 2006a). The overdensity of the Lagrangian volume, $\delta = \rho/\bar{\rho} - 1$, in the zoom simulation varies from $\delta \sim 0.1$ at about $z \sim 14$ to $\delta \sim 0.2$ at $z \sim 8$, where $\rho$ is the density of the Lagrangian volume. This volume is defined as all cells that contain a particle which eventually ends up in that respective halo, and $\bar{\rho}$ is the density of the full box. Thus, we report that in this HESTIA realization, the DM particles which eventually end up in the M31 and MW haloes, respectively, are constantly in an overdense region of the box volume. Although other studies have reported such a correlation between the redshift of reionization of a region and its overdensity (I. T. Iliev et al. 2006a, 2011; T. Dawoodbhoy et al. 2018; J. G. Sorce et al. 2022), studies by K. L. Dixon et al. (2018) showed that in some constrained realizations the LG haloes might reside in underdense regions. It is important to remember that these constrained simulations are mostly constrained on observational data of our local neighbourhood at $z = 0$ and thus there are multiple evolutionary pathways which the LG haloes might take (E. Wempe et al. 2025). Definitive results on this will require studying a wide range of constrained realizations.

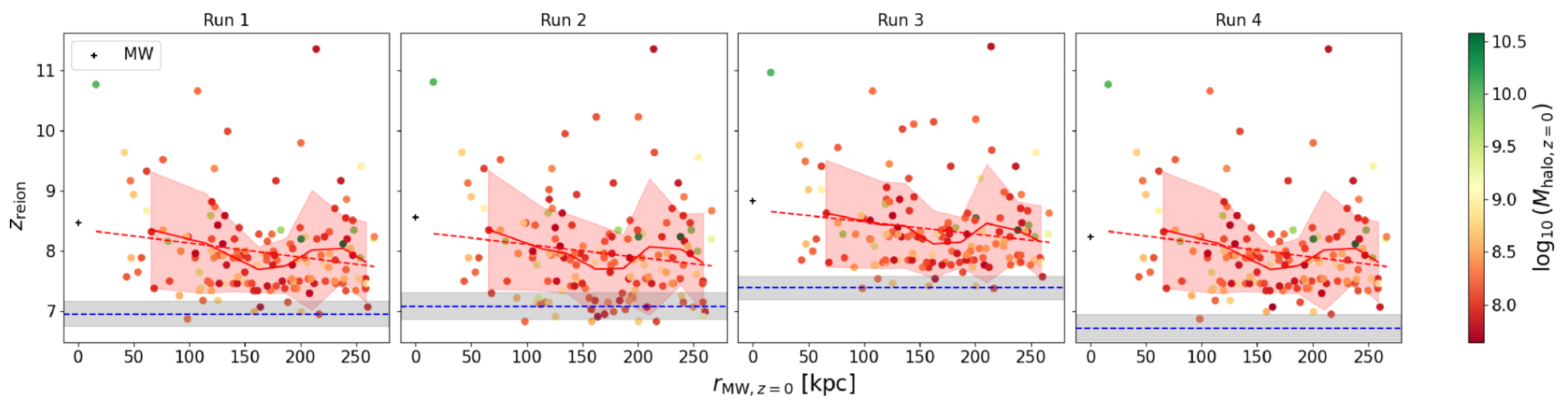

**Figure 6.** Top: The redshift of reionization ($z_{\rm reion}$) of the MW satellites and MW halo against the radial distance of the satellite from the MW halo at $z = 0$, where the satellite markers are coloured depending on the mass of the satellite at $z = 0$. The solid red line shows the binned average of $z_{\rm reion}$, and the red shaded region is the $1\sigma$ dispersion, with the dashed red line showing the linear best fit.

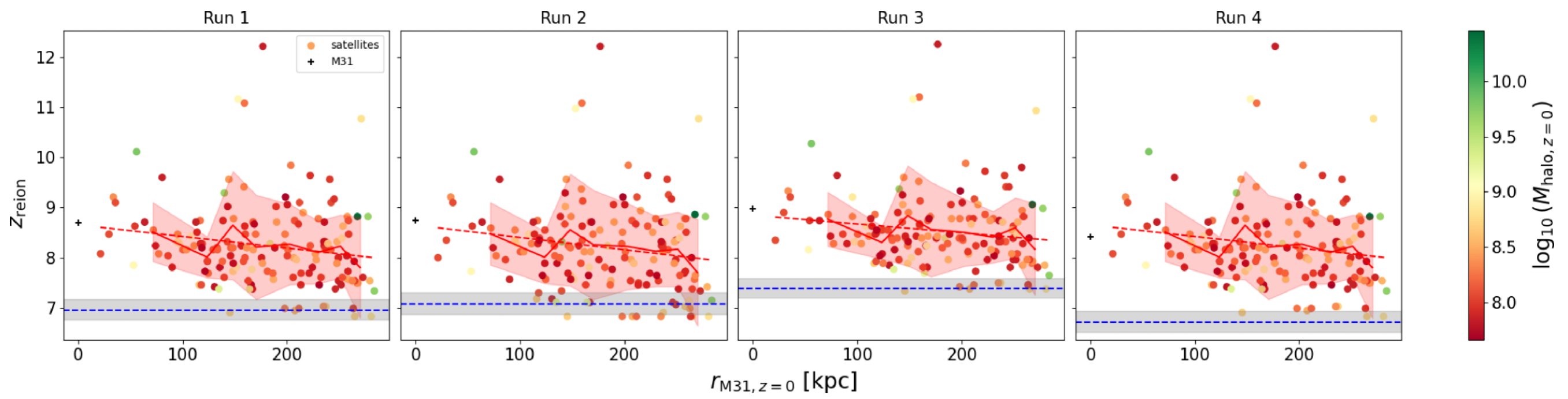

**Figure 7.** Same as Fig. 6, but for M31 satellites.

### 3.3 Satellite galaxies

In this analysis, we also look at the satellite dwarfs defined to be substructures within a virial radius of the MW and M31. We only look at well-resolved satellites with at least 20 DM particles identified at $z = 0$ using AHF. At $z = 0$, we find that the:

(i) MW has 150 satellites at $z = 0$ with a mass range of $10^{7.64}$–$10^{10.58}$ $M_\odot$;

(ii) M31 has 155 satellites at $z = 0$ with a mass range of $10^{7.69}$–$10^{10.46}$ $M_\odot$.

#### *3.3.1 Reionization history of satellites*

To understand how reionization progressed within the satellite galaxies of the MW and M31 haloes, in Figs 6 and 7, we plot the reionization redshift ($z_{\rm reion}$) against their distance away from the MW and M31 haloes, respectively, at $z = 0$. The scatter points are coloured based on the satellite halo mass at $z = 0$. The redshift of reionization of the satellites is solely calculated from the RT simulations run on the zoom simulations. We also plot the redshift at which the full box is predicted to reach a 50 per cent mass-weighted ionized fraction, as well as a shaded region at which the full box is predicted to have 40–60 per cent ionized fraction, based on the RT simulations run on the subgrid-coupled DMO simulations.

The Spearman rank correlation coefficient between the $z_{\rm reion}$ and $r_{\rm MW,z=0}$ for Run1_zoom, Run2_zoom, Run3_zoom, and Run4_zoom was found to be $-0.123$, $-0.122$, $-0.116$, and $-0.123$. This negative correlation implies that satellites that end up closer to the MW halo by $z = 0$ have 50 per cent of their material being reionized earlier than those that reside further away. This could be suggestive of an inside-out reionization scenario in the case of the MW satellites, whereby the MW halo is driving the reionization of its own satellites. However, this correlation is very weak and significant scatter can be observed in the $z_{\rm reion}$ for a given $r_{\rm MW,z=0}$. This scatter is caused due to the larger satellites being able to also contribute to the reionization of the region. Thus, despite reionization being mainly driven by the central halo (in this case the MW) and carried out mostly externally for the low-mass haloes, the larger satellites are able to reionize themselves as their progenitors would have accreted enough mass to produce ionizing photons by the time they reionize.

Similarly, the Spearman rank coefficients were calculated in the case of the M31 satellites, where we obtain coefficients of $-0.210$, $-0.193$, $-0.190$, and $-0.210$ for Run1_zoom, Run2_zoom, Run3_zoom, and Run4_zoom, respectively. We note a slightly stronger correlation between $z_{\rm reion}$ and $r_{\rm M31,z=0}$ com-pared to the MW satellites, as well as less scatter in the reionization times of satellites found at approximately the same present-day distance from the M31 halo. These correlations have also been noted in previous studies of reionization in the context of the LG (P. Ocvirk et al. 2014; K. L. Dixon et al. 2018). In particular, K. L. Dixon et al. (2018) also found smaller scatter in the values of $z_{\rm reion}$ for the M31 satellites when compared to the MW satellites, which hints that such a result might be more than just a coincidence and might be attributed to the constraints imposed to generate a realistic LG.

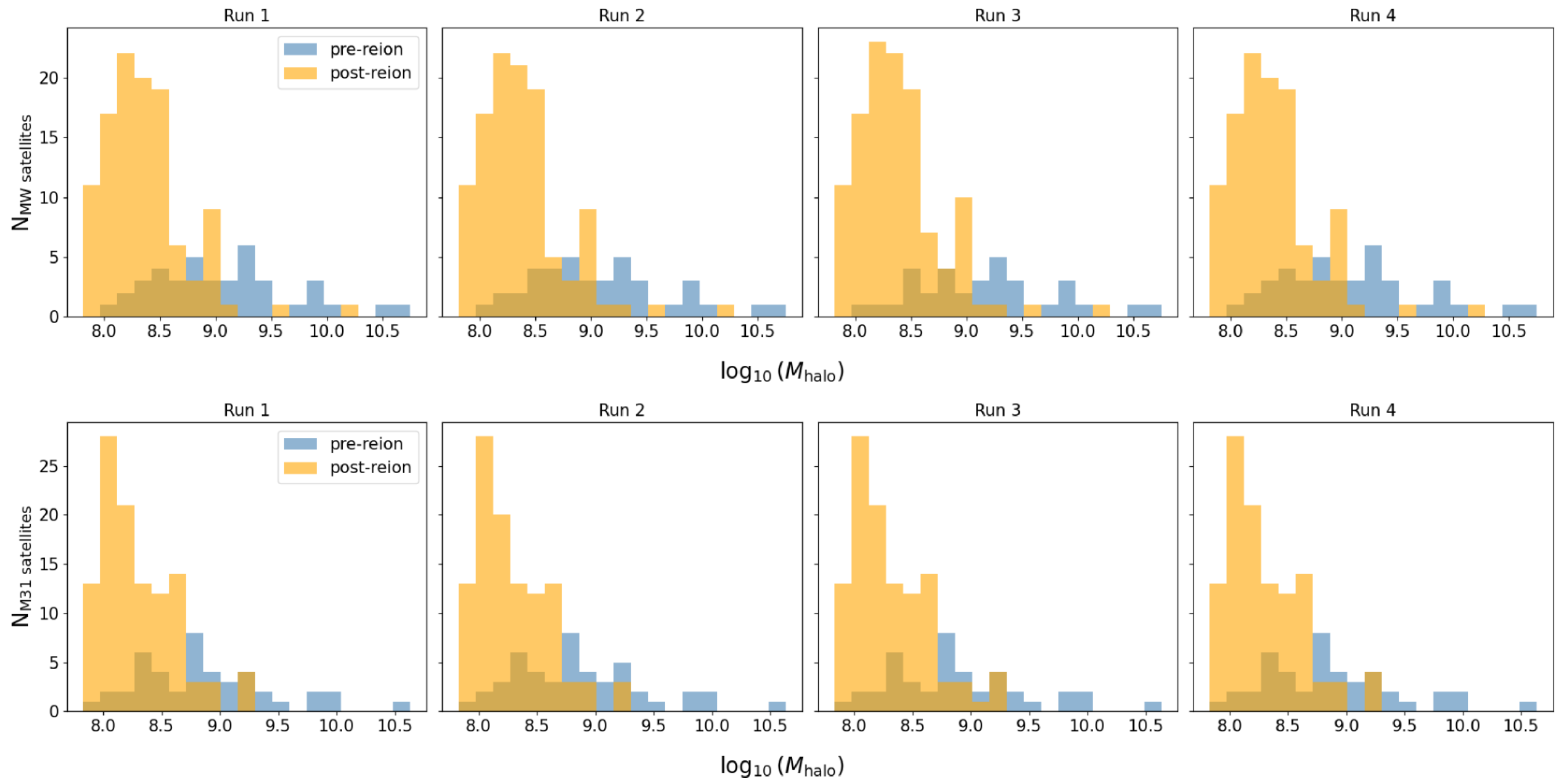

**Figure 8.** Top: Histograms of the present-day halo masses of MW satellites, split by formation history. Satellites that formed before their material was reionized are shown in blue, while those that formed after reionization are shown in orange. Results are shown for four different RT models. Bottom: Same as the top panel, but for satellites of M31.

Another observation from Figs 6 and 7 is that almost all the satellite galaxies have their material, which eventually ends up in their haloes, reionized much earlier than the predicted time for when the full box should reach 50 per cent ionized (obtained from the RT simulations run on the subgrid-coupled DMO simulations). This could be due to the fact that the material which eventually ends up in the LG continuously resides in overdense regions, which are expected to reionize earlier as noted in the previous section; furthermore, such satellites could in turn also be sources themselves and thus ionize the region around them.

Comparing the results between Run 1 and Run 2, no significant difference can be observed, further showing that the redshift of reionization of the full box is a more significant factor than the feedback model imposed. Moreover, when we compare Run 3 and Run 4 results, we mainly only observe a relative vertical shift between the two plots, which is caused due to the harsher feedback imposed, in turn delaying the overall reionization process, whilst conserving the overall shape of reionization.

### *3.3.2 Oldest stellar populations*

To establish a connection between the reionization history and the observable properties of the LG satellites, we classify the simulated satellites of the MW and M31 according to their assembly relative to the EoR. We define pre-reionization satellites as those which first appear in the merger tree at a redshift $z_{\rm form} > z_{\rm reion}$, while post-reionization satellites are those for which $z_{\rm form} < z_{\rm reion}$.

While this definition of $z_{\rm form}$ is inherently linked to the simulation's mass resolution, it serves as a proxy for whether a halo was present to initiate star formation prior to the suppression of gas inflow. Our simulation resolution captures any atomic cooling halo. Consequently, haloes appearing before reionization inhabit a neutral IGM where gas can cool and form stars unhindered. In contrast, those appearing only after reionization must form from an ionized IGM where the Jeans mass is significantly elevated, effectively preventing them from initiating star formation during their early growth. Crucially, because these resolved systems are above the atomic cooling limit, they are not subject to complete photoevaporation and can survive as subhaloes to $z = 0$. Thus, $z_{\rm form}$ distinguishes between satellites that could have hosted an early, unsuppressed stellar population and those whose star formation was delayed or prevented by the ionization of their environment.

We then compare the present-day ($z = 0$) halo mass distributions of these two populations, as shown in Fig. 8. Using the KS (Kolmogorov–Smirnov) test, we evaluate the null hypothesis that pre- and post-reionization halo mass distributions are drawn from the same parent population. Across all four RT model variants for both MW and M31 systems, we obtain $p < 0.001$, allowing us to reject the null hypothesis with high confidence. Thus, the data indicate that satellites hosting the oldest stellar populations preferentially inhabit more massive DM haloes at $z = 0$.

The mass separation observed between pre- and post-reionization satellites is not a trivial consequence of the correlation between halo mass and collapse redshift. The $z = 0$ mass of a satellite is the integrated result of its entire evolutionary history, including stochastic orbital evolution and environmental effects such as tidal stripping and dynamical friction. Since more massive subhaloes are often more susceptible to these disruptive forces (A. V. Kravtsov, O. Y. Gnedin & A. A. Klypin 2004; T. Sawala et al. 2016), the earliest forming systems are not guaranteed to remain the most massive survivors today. Our findings indicate that within a realistic LG context, the signature of early assembly remains a primary determinant of the present-day satellite population, despite the complex non-linear evolution of subhaloes.

## 4 DISCUSSION AND CONCLUSIONS

In this work, we have introduced and validated a computationally efficient pipeline for studying LG reionization: first calibrating ionizing source models on a coarse $1024^3$ DMO run coupled with a subgrid collapse fraction model, then applying those calibrated source prescriptions in high-resolution zoom simulations.

We note that an alternative and more conservative boundary treatment has been developed by R. Kannan et al. (2025), where high-cadence radiation maps from the parent box are interpolated in space and time and imposed outside the high-resolution region so that inflowing radiation is explicitly propagated into the zoom with the RHD (radiation-hydrodynamics) solver. This approach can capture situations in which a strong external ionizing front reionizes parts of the zoom region outside-in at early times. In the present HESTIA study, we instead validated the omission of a time-dependent external radiation map by first running a coarse full-box RT calibration, confirming an inside-out topology for the LG realization, and then rerunning high-resolution zooms without an imposed parent radiation field; this choice therefore minimizes computational cost while remaining appropriate for our LG reionization scenario. We acknowledge the THESAN-ZOOM boundary-mapping technique as a robust means to include large-scale, patchy radiation when (and only when) a zoom target lies in the path of strong external fluxes or when parent and zoom galaxy formation physics differ significantly.

We summarize our findings below:

(i) *An inexpensive, yet accurate method for RT zooms*. By benchmarking against the full-box DMO + subgrid runs, we show that zoom simulations, when properly calibrated, reproduce the MW and M31 reionization histories with minimal bias. This inside-out confirmation in a single HESTIA realization demonstrates that zoom RT simulations can be an effective approach for future LG studies.

(ii) *Inside-out reionization in all models*. In four source/feedback scenarios (two matched in global history but differing in suppression strength, and two sharing efficiencies but differing in suppression), the MW and M31 always reionize their surroundings from the inside-out. Large-scale external fronts play a negligible role even in our latest reionizing models, confirming that internal sources dominate LG reionization.

(iii) *LG haloes reionize earlier than the cosmic mean*. In every run, the mass-weighted reionization redshifts of the MW ($z_{\rm reion} \approx$ 8.7–9.2) and M31 ($z_{\rm reion} \approx$ 8.8–9.3) exceed the global 50 per cent point ($z_{\rm reion} \approx$ 7.1–7.7). We attribute this advance to the over-dense Lagrangian environments of the progenitors. Similar early reionization has been noted even in unconstrained scenarios (S. M. Weinmann et al. 2007; K. L. Dixon et al. 2018), motivating a larger sample of HESTIA realizations to pin down environmental effects.

(iv) *Older population stars in more massive satellites*. We find a correlation between the reionization time and the present-day mass of a satellite, where we find that satellites that formed prior to the time they underwent reionization tend to have larger masses. Such a correlation predicts that more massive satellites are more likely to harbour the oldest stars.

Despite exploring a broad parameter space in source efficiencies and suppression models, we find that the LG reionization morphology and timing remain remarkably stable. This underscores both the power of constrained zoom-in techniques and the dominance of internal processes in the LG's reionization. Future work will aim to quantify the environmental dependence of these results by applying the same calibrated pipeline to the full suite of HESTIA constrained realizations. This will allow us to (i) assess the variance in MW/M31 $z_{\rm reion}$ arising from different large-scale environments, and (ii) refine predictions for satellite quenching that can be tested against upcoming deep photometric and spectroscopic surveys.

## ACKNOWLEDGEMENTS

DA's work was supported by the Science and Technology Facilities Council (grant number ST/W006839/1) through the DISCnet Centre for Doctoral Training. Some of the analysis was done on the Apollo2 and ARTEMIS clusters at the University of Sussex. The authors gratefully acknowledge the Gauss Centre for Supercomputing e.V. (www.gauss-centre.eu) for funding this project by providing computing time through the John von Neumann Institute for Computing (NIC) on the GCS Supercomputer JUWELS at Jülich Supercomputing Centre (JSC). LC was supported by STFC consolidated grant ST/X000982/1.

## DATA AVAILABILITY

The data used in this paper are available for any reasonable requests to the corresponding author.

This paper has been typeset from a TEX/LATEX file prepared by the author.